\documentclass[a4paper,12pt]{article}
\usepackage[utf8]{inputenc}
\usepackage{graphicx}
\usepackage{authblk}
\usepackage{cite}
\usepackage{amsmath}
\title{Comparison of Spatial Entanglement between dissociated atom and ion : molecular ion photo-dissociated by sequential two-photon absorption and correlated two-photon absorption}

\author[1,*]{Krishna Rai Dastidar} 
\affil[1]{Indian Association for the Cultivation of Science, Kolkata-700032, INDIA}

\date{}

\begin{document}

\maketitle 
* spkrd@iacs.res.in

\begin{abstract}

We have studied the fidelity of spatial entanglement  between dissociated atom and ion from two photon dissociation of hydrogen  molecular ion ${H_2}^+$.
Two processes for two photon dissociation of molecular ion have been considered (i) sequential two photon (STP) absorption and (ii) correlated two photon (CTP) absorption by the molecular ion. We compared the results for fidelity of spatial entanglement (FSE) between dissociated atom and ion for these two types of dissociation. 
In a previous study in our group \cite{TKRD1998} 
we have shown that when an atom interact with nonlocal mode of electromagnetic field (which has been derived field theoretically), simultaneous phase-correlated two photon absorption by the atom occurs  within a very short time $\delta{t}<<{1\over{\omega}}$, where $\omega$ is the laser frequency, and the rate of this correlated two photon absorption is linear in intensity. This will happen when the photon flux in the interaction region is high i.e. laser intensity is higher than $10^{10} W/cm^2$. 
In this work we have extended this formalism to study the correlated two photon dissociation of molecular ion and to explore the effect of correlation between two simultaneous photo-absorption processes on the fidelity for spatial entanglement of dissociated atom and ion. Dependence of fidelity for spatial entanglement on the photon frequency has been studied for both the STP and CTP dissociation, to  show that photon frequency can be used as control parameter to achieve maximum fidelity. Saturation of fidelity with the increase in time at which fidelity was calculated shows the robustness of the processes considered here. A scheme for detection of spatial entanglement between ion and atom has been suggested.

\end{abstract}

\section{Introduction}
It is well known that nonlocal entanglement between pair-correlated particles leads to the fact that measurement on one particle can instantly affect the other, regardless of their distance. Quantum entanglement between two or more subsystems of a composite system plays a key role in quantum information processing and quantum computation \cite{Jurgen2002}. There are many demonstrations of such entanglement in photons and atomic internal states \cite{Tichy2011}. Entanglement between indistinguishable particles \cite{Eckert2002, Gian2004},  in fermionic system \cite{John2001, Zander2010} and in Bose Einstein Condensates (BECs) \cite{Pu2000,  Sorensen2001, You2003} has been studied. Attempts were made to prepare and study entanglement in dissociated atoms from molecules where molecules are excited to a dissociating state by collisions, by Feshbach resonance in BECs or by photo excitation \cite{Athreya2026, Tanabe2010, Esquivel2011, Gonis2014, Dochain2023, Torizuka2025, Eckart2023}. However
studies on momentum and position entanglement in two particles requires more attention for theoretical studies and experimental investigations.  In a theoretical simulation spatial correlation between two atoms from dissociation of molecules in molecular BEC was done \cite{Savage2010}. Generation and detection of high-fidelity spatial entanglement in two atoms prepared from molecular BEC via Feshbach resonance was studied using integrated atom optics \cite{Zhao2007}. 

In this work we will study on generation and detection of high-fidelity spatial entanglement between atom and ion obtained from two-photon dissociation of simplest diatomic molecule, Hydrogen molecular ion (${H_2}^+$). In a previous work in our group \cite{TKRD1998}, we proposed a model of nonlocal quantum electrodynamics which was derived from the first principle field theoretically. It was 
shown that electromagnetic field can be expressed as a sum of two modes local and non-local. For two photon excitation the intensity dependence of rate of transition  can be written as $R= \alpha I + \beta I^2$, where I is the laser intensity which means that it deviates from usual quadratic intensity dependence and the contribution from the quadratic intensity dependence part diminishes with increase in intensity compared to that from interaction with nonlocal mode of electromagnetic field.
This feature was observed in an experiment to explore squeezed vacuum states \cite{Georgiade1995}. By using this nonlocal model of electromagnetic field it has been shown that the superposition of dressed states gives rise to the same form as squeezed vacuum state. 
It has also been shown that the contribution from the nonlocal mode of electromagnetic field (which is linear in intensity) in the rate of two-photon ionization can change the slope of the rate curve with increase in laser intensity \cite{KRD1988, KRD2010} .  This nonlocal model of electromagnetic field was applied to explore various aspects of quantum physics \cite{TKRD1999}. 

In two photon absorption by a molecule usual concept is that it will be absorbed sequentially which is the case for interaction with local mode of electromagnetic field and it will give rise to two-photon absorption rate which is  proportional to the square of laser intensity. However when the laser intensity is high enough ($> 10^{10} W/cm^2$) interaction with nonlocal mode of electromagnetic field will be effective. For high laser intensity photon flux in the interaction region will be high and two photons will interact with the molecule simultaneously within a very short time ($\delta{t}<<{1\over{\omega}}$, where $\omega$ is the laser frequency). Hence  two photons will be phase correlated leading to two photo-absorption processes to be correlated. Thus simultaneous absorption of two correlated photons gives rise to two-photon transition amplitude which varies as square root of laser intensity \cite{TKRD1998}.

The aim of this work is to explore the effect of the correlated two photon absorption (simultaneous) on the 
fidelity of spatial entanglement of  dissociated atom and ion obtained from the dissociation of  molecular ion and to compare the results with those obtained from the sequential two photon dissociation due to interaction of molecular ion with the local mode of electromagnetic field.

This work has been arranged as follows:
In section 2 (Theory), there are three subsections describing theory for (2.1) Sequential two-photon (STP) dissociation, (2.2) correlated two-photon (CTP) dissociation  and the (2.3) theory for fidelity of spatial entanglement. In section 3 (Numerical calculation), method of calculation of wavepackets on the excited states has been described. Analytical method for calculation of electronic dipole transition moment in case of correlated two-photon dissociation has been given. In section 4 (Results and Discussion) results were presented in two subsections 4.1 (Dependence on photon frequency) and 4.2 (Dependence on time). In section 5, Conclusion was drawn.

\section {Theory}
In this work we have investigated the fidelity of spatial entanglement of atom and ion in two-photon dissociation of hydrogen molecular ion ${H_2}^+$. We compared the degree of entanglement of H-atom and proton obtained by sequential two-photon (STP) dissociation \cite{Bhatt2003} and by correlated two-photon (CTP) dissociation \cite{TKRD1998} of hydrogen molecular ion by using intense femtosecond laser. In sequential two-photon dissociation molecular ion is excited from the ground ro-vibrational level of ground  $1s \sigma_g$ state to the first excited  $2p\sigma_u$ state and then by the second photon it is excited from $2p\sigma_u$ state to the dissociating $2s \sigma_g$ state and it dissociates to  atom and ion of equal energy. Here dissociation occurs due to the interaction of molecular ion with the local mode of electromagnetic field. 

To calculate the correlated two photon excitation we have considered interaction with nonlocal mode of electromagnetic field \cite{TKRD1998} in which it is considered that the flux of photons are high enough so that the molecule faces two phase correlated photons within a very short time interval $\delta_t<< {1 \over {\omega}}$  ( where $\omega$ is the photon frequency ) and the two photons are absorbed simultaneously leading to  molecular excitation coherently from the ground  $1s \sigma_g$ state to the dissociating $2s \sigma_g$ state. Unlike sequential two-photon excitation where rate is proportional to the square of the laser intensity, the rate of correlated two-photon excitation is linear in intensity.  
For the generation of final wavepacket on the dissociating state ($2s \sigma_g$) we have used wavepacket formalism \cite{Bhatt2003}, both for sequential two-photon dissociation and correlated two-photon dissociation of ${H_2}^+$ ion .

\subsection{Wavepacket generation on the final dissociating state in sequential two-photon (STP) excitation :}
Here we have considered by the first photon (pulse) the molecular wavepacket in the ground ro-vibrational level of the ground $1s \sigma_g$ state at t=0 is photo-excited (dipole transition) to the excited state $2p \sigma_u$.  The amplitude of the wavepacket  on the excited state at time t=T (where T is the pulse duration) and at internuclear separation R is given as:
\begin{equation}
\begin{split}
(i\hbar) X_e (R,T)  =  \int_{0}^{T}  \exp ({-i H_e (T -t_1) \over \hbar}) [D_{eg}(R)] \epsilon (t_1) \\
\times{\exp({-i H_g t_1 \over \hbar}) X_{gv}(R,0) d t_1} 
\end{split}
\end{equation}

Where $X_{gv} (R,0)$ is the wave function of the vibrational level ‘v’ of the ground state 'g' at the time t=0. $H_g$ and $H_e$ are the total Hamiltonian of the ground state ‘g’ and the excited state ‘e’ respectively. $D_{ge}$ is the electronic dipole transition moment for transition from the ground state ‘g’ to the excited state ‘e’.
Then by the second photon the wavepacket on first excited state is further excited to the dissociating state $2s \sigma_g$.
The wave packet on the final dissociating state is given as:
\begin{equation}
\begin{split}
(i\hbar) X_f (R,T)  =  \int_{T+\tau}^{T'}  \exp ({-i H_f (T' -t_2) \over \hbar}) [D_{fe}(R)] \epsilon (t_2) \\
\times{\exp({-i H_e t_2 \over \hbar}) X_e(R,T+\tau) d t_2} 
\end{split}
\end{equation}

Where $X_e(R,T+\tau)$ is the propagated wave packet on the state ‘e’. $D_{fe}$ is the electronic dipole transition moment from the state ‘e’ to the state ‘f’ and $H_f$ is the total Hamiltonian of the state 'f'. For this transition electric field is on for the duration $T+\tau$ to T'. $\epsilon ( t_i)$ is the electric field at time $t_i$ given by 
\begin{equation}
 \epsilon(t_i) = g(t_i) \exp (i\omega t_i)
\end{equation}

where $g(t_i)$ is the temporal profile of the pulse and $\omega$ is the carrier frequency. For details see \cite{Bhatt2003}.

\subsection{Wavepacket generation on the final dissociating state in correlated two-photon (CTP) exciation :}

In this case molecular ion is excited from the ground ro-vibrational level of the ground $1s \sigma_g$ state directly to the dissociating $2s \sigma_g$ state by phase correlated two photon absorption. Here the correlated two-photon dipole transition occurs due to the interaction of electron with the nonlocal mode of electromagnetic field. 

Previously it has been shown \cite{TKRD1998} that in the interaction with the nonlocal mode of electromagnetic field correlated two-photon excitation of atoms (single electron or two electrons) can occur where the electronic dipole transition amplitude is proportional to square root of laser intensity and obey the energy conservation rule i.e. $ E_f=E_i+2\hbar \omega$. Here $E_f$ and $E_i$ are the energy of the excited and the initial state respectively and $\omega$ is the carrier frequency of the field. 

In this work we have extended the formalism for correlated two photon excitation of atoms to the correlated two photon excitation of molecules with single electron. For molecules, the correlated two-photon dipole transition occurs from the initial state to the excited state of same symmetry; the total wave functions of these states are given as $ \Psi_i (\vec{r}’, \vec{R}’)$ and $\Psi_f (\vec{r}, \vec{R})$, respectively. Here $\vec{r}$ and $\vec{r}’$ are the electronic coordinates while $\vec{R}$ and $\vec{R}’$ are the nuclear coordinates. Using Born-Oppenheimer approximation these two wave functions are given as the product of electronic wave functions and nuclear wave functions as follows:
\begin{equation}
 \Psi_i (\vec{r}', \vec{R}')= \phi_i (\vec{r}';R') \psi_i (\vec{R}') 
 \end{equation}
  and
 \begin{equation}
\Psi_f (\vec{r}, \vec{R})= \phi_f (\vec{r};R) \psi_f (\vec{R}) 
\end{equation}
where the electronic wavefunction $\phi_i (\vec{r}';R')$ and $\phi_f (\vec{r};R)$ are given as follows:
\begin{equation}
 \phi_i (\vec{r}';R')= N_{ia} {R_i(r')\over {r'}} Y_{{l_i}{m_i}}(\hat {r}') 
 \end{equation}
 and
 \begin{equation}
\phi_f (\vec{r};R)= N_{fa} {R_f(r)\over {r}} Y_{{l_f}{m_f}}(\hat {r})
\end{equation}
respectively. Here $N_{ia}$ and $N_{fa}$ are the normalizing constant for the inintial and final electronic wavefunctions respectively. $R_i(r')$ and $R_f(r)$ are the radial electronic wavefunctions for the initial and final states respectively. l's and m's are the corresponding orbital and azimuthal quantum numbers.
The Nuclear wavefunctions are given as
\begin{equation}
 \psi_i (\vec{R'}) = N_{im} X_i(R') Y_{{L_i}{M_i}} (\hat {R}')
\end{equation}
and
\begin{equation}
 \psi_f (\vec{R}) = N_{fm} X_f(R) Y_{{L_f}{M_f}} (\hat {R}) 
 \end{equation}
 respectively. Here $N_{im}$, $N_{fm}$ are the normalizing constants and $X_i(R')$, $X_f(R)$
are the radial nuclear wavefunctions of the initial and final state respectively. L's and M's are the  rotational and magnetic quantum numbers respectively.

Hence the correlated two-photon electronic dipole transition amplitude is given as:
\begin{equation}
 T_{if} (R,R') =  {i \sqrt{2\pi I} \over c}N_a <  {\phi_f (\vec{r};R) }| { r . {\hat{\epsilon}(\hat {r}, \hat {r}' )} \vert { \phi_i (\vec {r'};R')}}>
\end{equation}
Where 'I' is the intensity of laser field, 'c' is the velocity of light and $N_a=N_{ia} N_{fa}$.
The non-local electromagnetic field $\hat{\epsilon}(\hat {r}, \hat {r'})$ was previously given as
\begin{equation}
 \hat {\epsilon}(\hat {r}, \hat{r}') = (\hat {r}+\hat {r}') \sum_{n=0}^{\infty} a_n P_n (\hat {r}, \hat {r}')
 \end{equation}
(here symmetry demands that the coefficients of r and r’ are identical).
 Hence putting equation (11) into equation (10), two-photon electronic dipole transition amplitude can be written as:
\begin{equation}
 T_{if} (R,R') = {i \sqrt{2\pi I} \over c} \textbf{R} (I_1+I_2)
 \end{equation}
Where  \textbf{R} is the radial integral and I's are the angular integrals.  Hence from above equation (12), we find that the correlated two-photon transition amplitude is proportional to square root of intensity. Elementary analysis shows that 
\begin{equation}
 \textbf{R}=N_a {\int_{0}^{\infty} R_i(r') {r'}^2 dr'} {\int_{0}^{\infty} R_f(r)r dr}
\end{equation}
\begin{equation}
 I_1= \sum_{n=0}^{\infty} {{4\pi a_n} \over 2n+1} \sum_{\nu=-n}^{n} \int {Y_{l_fm_f}}^*(\hat {r}) Y_{n\nu} (\hat {r})d{\hat{r}} \int {Y_{n\nu}}^*(\hat {r}') Y_{l_im_i}(\hat {r'}) d {\hat{r'}}
\end{equation}
\begin{equation}
\begin{split}
 I_2= \sum_{n=0}^{\infty} {{16{\pi}^2 a_n} \over 3(2n+1)} \sum_{\nu=-n}^{n} \sum_{m=-1}^{m} \int {Y_{l_fm_f}}^*(\hat {r}) Y_{1m} (\hat {r}) Y_{n\nu} (\hat {r}) d{\hat{r}} \\
 \times{\int {Y_{1m}}^* (\hat {r}') {Y_{n\nu}}^*(\hat{ r}') Y_{l_im_i}(\hat {r}') d {\hat{r}'}} 
\end{split}
\end{equation}

The first angular integral gives the selection rule $l_f=l_i$ and the second integral gives the selection rule $l_f=l_i, l_i \pm {2}$ (see Appendix C of \cite{TKRD1998}).
In this case we have considered $l_f=l_i=0$ and $m_f=m_i=0$.
Hence for the correlated two-photon transition from ground to final dissociating $2s\sigma_g$ state, the wavepacket on the dissociating  state is given as:
\begin{equation}
\begin{split}
(i\hbar) W_{if} (R,R',T)  = \int_{0}^{T}  \exp ({-i H_f (T -t_1) \over \hbar}) [D_{if}(R,R')] \epsilon (t_1) \\
\times {\exp({-i H_g t_1 \over \hbar}) X_{gv}(R,0) d t_1} 
\end{split}
\end{equation}

where $D_{if}(R,R') =\textbf{R} (I_1+I_2)$ is the dipole transition moment for transition from ground state to final dissociating state.
T is the duration of the pulse of amplitude $\epsilon(t_1)$, R and R' are the relative nuclear coordinates of atom and ion on excited and ground state respectively. $H_g$ and $H_f$ are the total Hamiltonian of the ground and dissociating excited state respectively. We have considered here excitation to the dissociatiing state occurs from the vibrational level $v_i=0$ with $L_i=0$ of the ground state and considering  there is no change in rotational quantum number during excitation, we put $L_i=L_f=0$ and hence $M_i=M_f=0$ in this calculation.  In this work we have analytically calculated $D_{if}(R,R')$ (see Appendix).

\subsection{Dynamics of dissociated atom and ion  and the Fidelity of spatial entanglement:}
The ${H_2}^+$ ion after being excited to the dissociating $2s\sigma_g$ state by two photon excitation from the ground $1s\sigma_g$ state will freely decay to entangled H-atom and proton. If one of them say atom leaves along right direction $a_1$ with momentum $\vec{k_a}$, the ion will move in the left direction $a_2$ with momentum $-\vec{k_a}$. But there will be probability for moving entangled atom and ion in many pair of correlated directions hence the probability for moving atom and ion in any specified direction will be low. This difficulty can be avoided using integrated atom optics (for review see \cite{Folman2002}) where atom and ion can be streamlined in one or two correlated directions by using atomic and ionic waveguides. In each side, after introducing a phase shift in one part by phase shifter, atomic/ionic waves can be splitted into two parts by using  beam splitter and then two parts will be reunited for interferometric detection by a detector. For atom and ion moving in the opposite direction can be detected in coincidence by  detectors placed in the opposite  directions. In the present case since the molecular ion dissociates as H atom and ${H^+}$ ion, for guiding the atom in one direction (1D) one will have to use neutral atom waveguide and for ion one can use RF-guide. Thus measuring the probability of atom and ion moving in same direction or in opposite directions one can find out the fidelity of spatial entanglement between atom and ion. 

In the case of selecting two channels a and b, first of all molecule is passed  through a beam splitter (BS), such that molecule will be splited in two parts towards the channels a and b.  Dissociated atoms (or atom and ion) in the channel 'a' will proceed in the opposite directions $a_1$ and $a_2$ with momenta $\vec{k_a}$ and $-\vec{k_a}$ respectively and those in the channel 'b' will proceed in the directions $b_1$ and $b_2$ with momenta $\vec{k_b}$ and $-\vec{k_b}$ respectively. The atoms (or atom and ion) moving in the $a_1$ and $b_1$ directions will be detected by a detector ($A_1$) placed on the right by introducing a phase $\phi_1$ in the path of $b_1$ and then passing through a beam splitter. Similarly the atoms moving in the directions $a_2$ and $b_2$ will be detected by a detector ($A_2$), placed opposite to $A_1$ after introducing a phase shift $\phi_2$ in the path of $b_2$ and then passing through a beam splitter for interferometric detection (for details of this scheme see \cite{Zhao2007}). For detection of FSE between two neutral atoms, integrated atom optics can be used where both the arms for detection contains neutral atom waveguide, phase shifter beam splitter and detector. But in this case for detection of FSE between atom and ion one of the two arms should be replaced by ionic waveguide, phase shifter, ion beam splitter and trapped ion detector.
  
  One dimensional motion of atoms and ions is controlled by 1D atomic/ionic waveguides. In this case one can obtain a path-entangled state as
\begin{equation}
 |\Phi>_{path}=\alpha|a_1 a_2> + \beta |b_1 b_2>
 \end{equation}
where ${\alpha}^2+{\beta}^2=1$ and $|a>$, $|b>$ are two orthogonal spatial states of atom/ion.
If the BS is perfect(50/50) then after dissociation one of the entangled state can be written as:
\begin{equation}
{\Phi}^->_{path}={1\over{\sqrt{2}}}{(|\Psi_{a_1 a_2}> - |\Psi_ {b_1 b_2}>)}
\end{equation}
where $|\Psi_{a_1 a_2}>$ and $|\Psi_ {b_1 b_2}>$ are the over all wavefunctions of the two branches. Since the channels 'a' and 'b' are identical we will consider here only one channel 'a'. 
After excitation of the molecular ion to the dissociating state the generated wavepacket will represent the entangled ion and atom in their relative coordinate R where R can be written in the position coordinates of atom ($R_1$) and ion ($R_2$) in one dimension (1D) as $R=R_1-R_2$. To find out the probability of atom and ion moving in the same direction or in the opposite direction one will require wavepacket in position coordinates of atom and ion which can be derived from the wavepacket in relative coordinates  of atom and ion as follows:
\begin{equation} 
 |Z_f(R_1,R_2,R',T)|^2 = | W_{if}(R,R’,T)|^2 \times  |\Phi(X,T)|^2 
\end{equation}
where $W_{if}(R,R’,T)$ is the wavepacket generated on the dissociating state (in relative coordinate) by absorbing two correlated photons (see equation (16))
and
$\Phi(X,T)$  is the wavepacket for the centre of mass (c.m.) motion. X is the position coordinate of the centre of mass. The c.m. wavepacket can be approximated as Gaussian functions as:
\begin{equation}
 |\Phi(X,T)|^2= {1 \over{ \sqrt{2\pi}\Delta {X_t}}} \exp {[-{X^2 \over {2(\Delta{X_t})^2}}]}
 \end {equation}
where 
\begin{equation}
\Delta{X_t} = \sqrt { (\Delta {X_0})^2 + {({i\hbar \over 4m\Delta{X_0}})^2}} 
\end{equation}
$\Delta{X_0}$ is the initial width of the gaussian wavepacket. 'm' is the reduced mass of the molecular ion. 

In case of dissociation by absorption of two sequential photons $W_{if}(R,R’,T)$ in (18) should be replaced by  $X_f (R,T)$ given in equation (2) and the wavepacket on the dissociating state in terms of position coordinates of atom and ion will be given as:
\begin{equation} 
 |S_f(R_1,R_2,T)|^2 = | X_f(R,T)|^2 \times  |\Phi(X,T)|^2 
 \end{equation}

To find out the fidelity for entanglement between dissociating atom and ion one will have to find out the second order correlation function at time T, 
for atom and ion moving in the same direction and moving in the opposite direction. The second order correlation function of atom and ion moving in the same direction $a_1$ for CTP absorption is given as 
\begin{equation}
{G^{(2)}_{a_1a_1}} = \int_{0}^{\infty} d{R_1} \int_{0}^{\infty} d{R_2}   |Z_f(R_1,R_2,R',T)|^2
\end{equation}
and that for atom and ion moving in the opposite directions $a_1$ and $a_2$ is given as
\begin{equation}
 {G^{(2)}_{a_1a_2}} = \int_{0}^{\infty} d{R_1} \int_{-\infty}^{0} d{R_2}   |Z_f(R_1,R_2,R',T)|^2
 \end{equation}

 Similarly to obtain second order correlation functions for STP absorption one will have to replace $|Z_f(R_1,R_2,R',T)|^2$ by $|S_f(R_1,R_2,T)|^2$ 
 in equations (23) and (24).

The probability for finding atom and ion in the same output ( i.e. moving in the same direction $a_1$)  or in the opposite output (i.e. moving in the opposite directions $a_1$ and $a_2$) is given as  ratio between two second order correlation functions as
\begin{equation}
 \kappa = {{G^{(2)}_{a_1a_1}} \over {G^{(2)}_{a_1a_2}}} 
\end{equation}
Hence the fidelity F can be written as
\begin{equation}
     F={1\over {1+\kappa} }
\end{equation}     
Therefore as $\kappa$ approaches zero, fidelity for spatial entanglement approaches unity leading to maximum spatial entanglement and as $\kappa$ becomes large fidelity becomes small resulting to minimum spatial entanglement. 

 \section {Numerical Calculation}
 For this calculation we have considered Hydrogen molecular ion (${H_2}^+$), which is a single electron molecule.
 For sequential  two photon dissociation from ground $1s\sigma_g$ to the dissociating $2s\sigma_g$ state via intermediate $2p\sigma_u$ state, all the electronic potential energy curves \cite{Sharp1971} and the dipole transition moments $D_{eg}$ and $D_{fe}$ were taken from the
literature (\cite{Bates1951, Bates1954, Ramaker1973}). 
For sequential two photon dissociation, we have  calculated the wavepackets on the intermediate $2p\sigma_u$ state and on the final dissociating state $2s\sigma_g$ using wavepacket formalism. The wave function for the ground vibrational state $X_{vg}$ was calculated using Fourier Grid Hamiltonian method \cite{Bhatt2003}.
\begin{figure}
\centering
\includegraphics[width=.60\textwidth]
{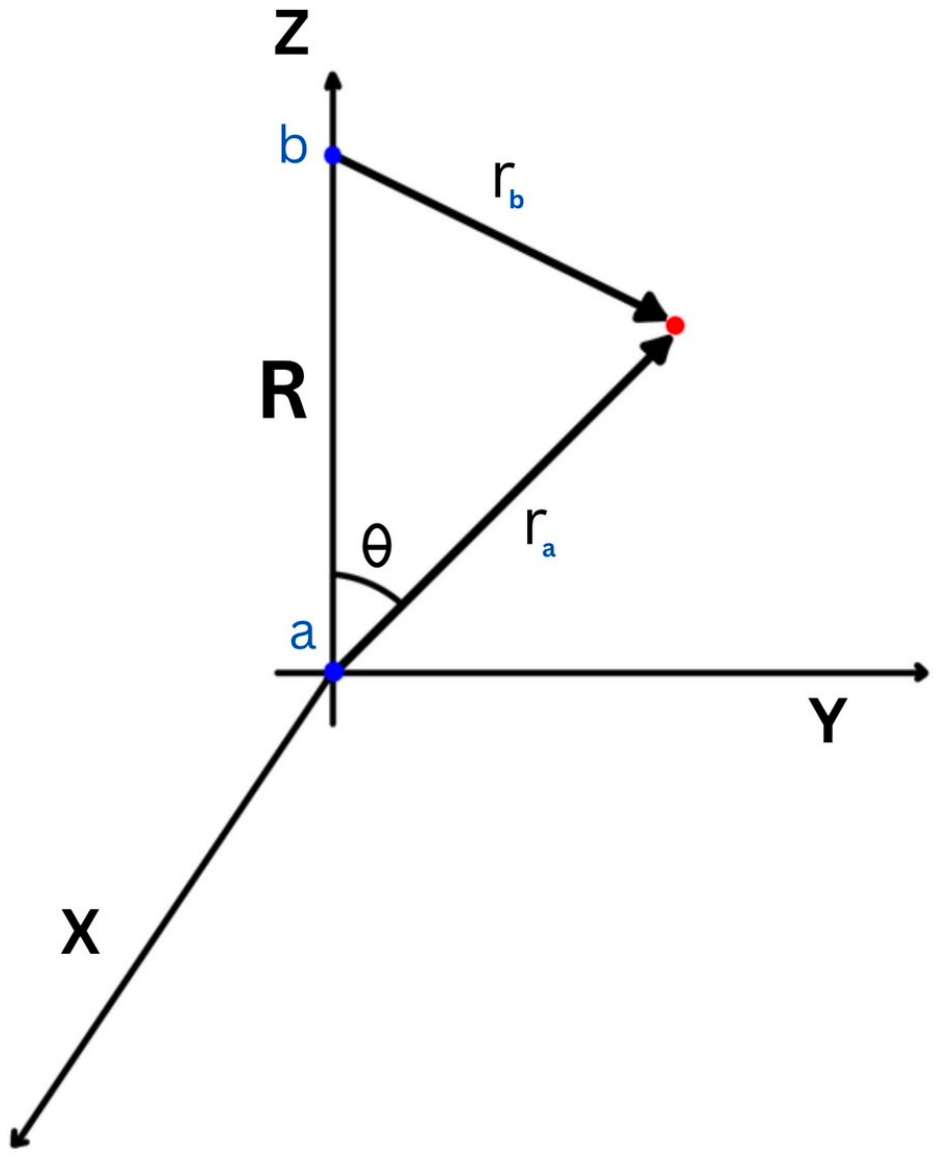}

{Fig1: Configuration for ${H_2}^+$ ion. R is the internuclear separation between the protons 'a' and 'b'. $r_a$ and $r_b$ are the vectors connecting  electron with neuclei 'a' and 'b' respectively. $\theta$ is the angle between R and $r_a$.} 
\end{figure}

For correlated two-photon transition we have calculated two-photon dipole transition amplitude analytically by using the analytical wavefunctions for the ground $1s\sigma_g$ and dissociating $2s\sigma_g$ state of ${H_2}^+$ ion. These wavefunctions were obtained by considering the configuration (Fig1) as given in literature \cite{Segre2020}. For details see Appendix.
The wavepacket on  the final dissociating state was calculated by using wavepacket formalism as mentioned above. 
In this calculation R'=2 a.u., the equilibrium
separation of the ground state.
To calculate the second order correlation functions for atom and ion moving in the same and opposite direction,
we considered the range of motion of atom and ion  $R_1$ and $R_2$ to vary from -20 a.u. to 20 a.u.
Total time for the propagation of wavepacket on the dissociating state has been considered to be 6700 a.u.. We have considered short femtosecond laser pulses, the duration of which is 6372 a.u. i.e. 
154.66 fs. Short pulses are used to maintain coherence in photo-dissociation \cite{Eckart2023}.  

\section{Results and Discussions}

In this work we have considered two schemes for two photon dissociation of single cold ${H_2}^+$
ion (production of this ion is within the scope of the present day facilities \cite{Holzapfel2025, Hudson2016, Julian2020}): 

(i) sequential two photon (STP) dissociation  due to interaction of molecular ion with local mode of electromagnetic field and 

(ii) correlated two photon (CTP) dissociation of molecular ion due to interaction with nonlocal mode of electromagnetic field as described in section 2 ('Theory'). 

The fidelity of spatial entanglement of H and $H^+$ was calculated for both the dissociation schemes (STP and CTP transitions) and compared the results.
 
 Close analysis of the results for fidelity of spatial entanglement of H and $H^+$ show that when the probability for finding  atom and ion in the same channel/detector decreases and/or probability for finding the atom and ion in two opposite detectors increases then the value of $\kappa$ decreases and the fidelity of spatial entanglement increases (see equations (25) and (26)).
Here the fidelity of spatial entanglement of H and proton has been plotted as a function of the initial width of the gaussian wavepacket (WGW) for the centre of mass (c.m.)  in units of 0.397 a.u., which is proportional to the full width at half maximum (FWHM) of the ground vibrational wavefunction of $1s\sigma_g$ state.  

In this work we explored two aspects of fidelity for spatial entanglement:

(i) dependence of fidelity of spatial entanglement on the frequency of photons for two photon dissociation and

(ii) Variation of fidelity with the choice of propagation time of the wavepacket on the dissociating state, at which the fidelity of spatial entanglement was calculated.  

\subsection{Dependence on Photon Frequency:}
To study the variation of fidelity of 
entanglement with the 
frequency of photons, three different frequencies were chosen 
corresponding to three different values of internuclear distance (R) on the dissociating  $2s\sigma_g$ state. The photon energies for transitions were calculated by subtracting the energy of the ground ro-vibrational level of $1s\sigma_g$ state from the energy of the dissociating $2s\sigma_g$ state at different internuclear separations R=2,3 and 4 a.u. and divided this energy difference by two. Hence the corresponding photon energies are 0.368267 a.u., 0.3059225 a.u. and 0.279422 a.u. respectively. We have calculated the fidelity of entanglement for these three frequencies as mentioned above at three different values of time 3000 a.u., 5000 a.u. and 6500 a.u. and  the results for three values of photon frequency at each value of time have been plotted in figures 2, 3 and 4 respectively. In all the figures calculated points have been connected by lines.

\begin{figure}
\centering
\includegraphics[width=.65\textwidth]{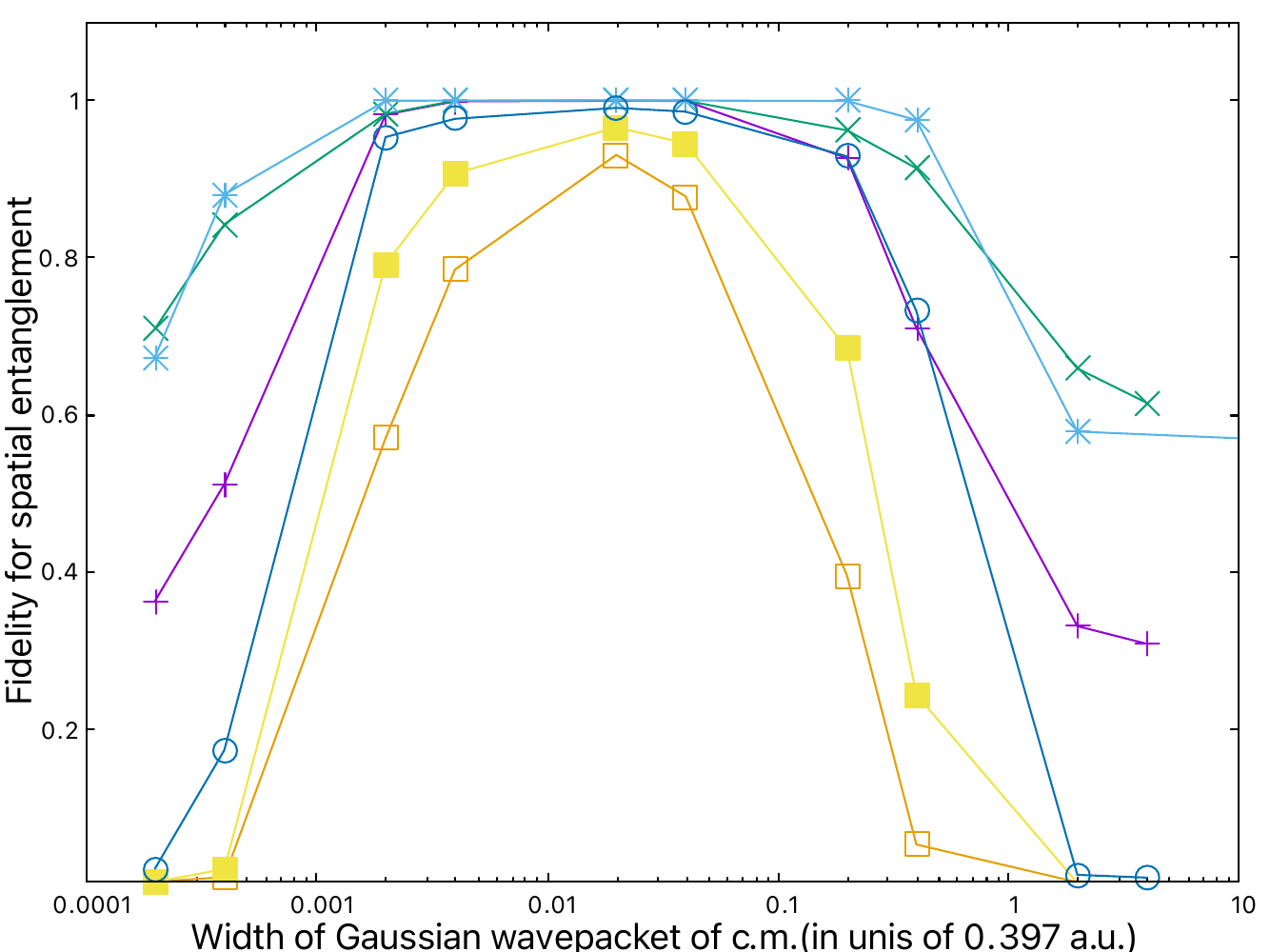}

{Fig2: Fidelity for spatial entanglement between dissociated atom and ion from (i) sequential two photon dissociation and (ii) correlated two photon dissociation of ${H_2}^+$ ion as a function of initial width of the gaussian wavepacket of centre of mass for three values of photon frequency at time=3000 a.u.
(i) For sequential two photon dissociation:
Open square - photon frequency = 0.368267 a.u.,
Filled square - photon frequency = 0.3059225 a.u.,
Open circle - photon frequency = 0.279422 a.u.
(ii) For correlated two photon dissociation:
Plus - Photon frequency = 0.368267 a.u.,
Cross - Photon frequency = 0.3059225 a.u.,
Star - Photon frequency = 0.279422 a.u.} 
\end{figure}

In each figure 2, 3 and 4, fidelity for spatial entanglement at three frequencies of transitions mentioned above has been plotted as a function of WGW, both for STP  and CTP dissociation for comparison.

\begin{figure}
\centering
\includegraphics[width=.65\textwidth]{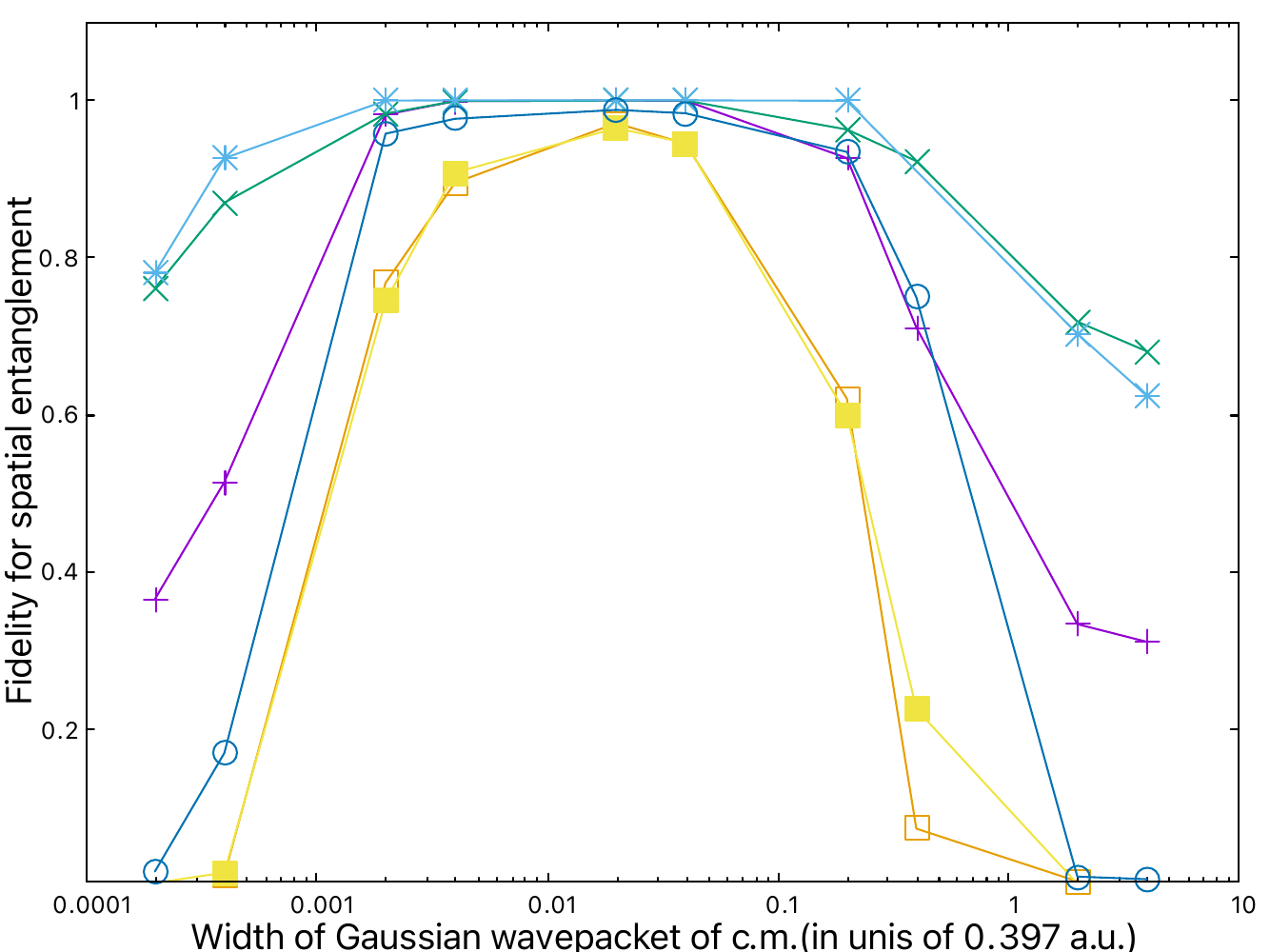}

 {Fig3: Fidelity for spatial entanglement between dissociated atom and ion from (i) sequential two photon dissociation and (ii) correlated two photon dissociation of ${H_2}^+$ ion as a function of initial width of the gaussian wavepacket of centre of mass for three values of photon frequency at time=5000 a.u. 
Both for (i) sequential two photon dissociation and
(ii) correlated two photon dissociation: Symbols and corresponding frequencies are the same as in Fig2}
\end{figure}

\begin{figure}
\centering
\includegraphics[width=.65\textwidth]{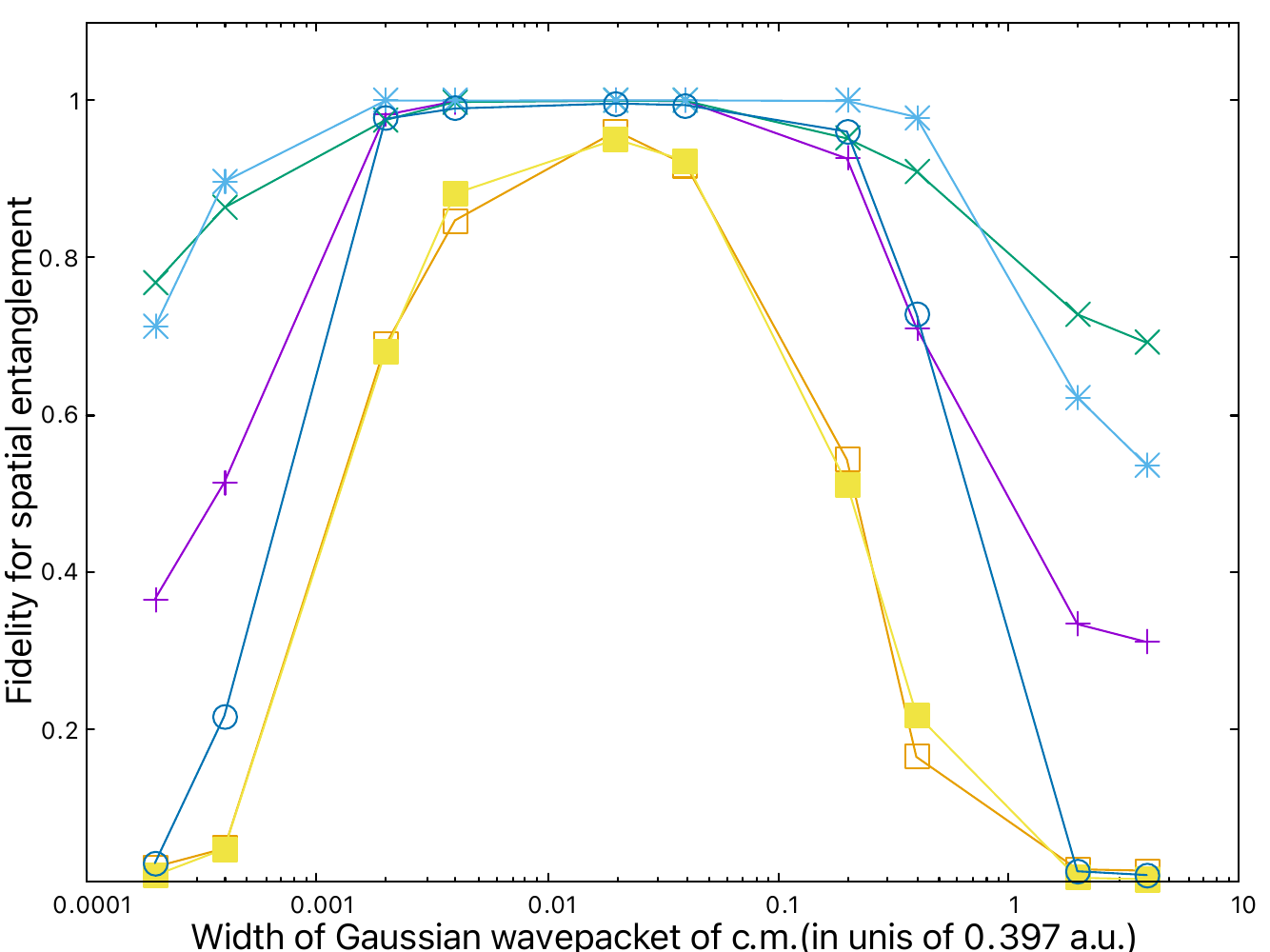}

 {Fig4: Fidelity for spatial entanglement between dissociated atom and ion from (i) sequential two photon dissociation and (ii) correlated two photon dissociation of ${H_2}^+$ ion as a function of initial width of the gaussian wavepacket of centre of mass for three values of photon frequency at time=6500 a.u. 
Both for (i) sequential two photon dissociation and
(ii) correlated two photon dissociation: Symbols and corresponding frequencies are the same as in Fig2.}
\end{figure}

  From figures 2,3 and 4 we note the following:
 
 (1) These figures show that in both the cases of STP and CTP dissociation, fidelity remains high for intermediate values of WGW and falls down to smaller values both for smaller and larger values of WGW. Hence for broad and narrow wavepackets for the centre of mass, probability of finding the atom and ion in the same detector becomes larger than that for finding the atom and ion in the opposite detectors, thus leading to smaller value of fidelity of spatial entanglement. This feature of fidelity of spatial entanglement has been reported previously \cite{Zhao2007}. 
 
(2) It is found that for the sequential two-photon (STP) absorption variation of FSE is  sharp i.e. it increases sharply from very small values and also falls down sharply to smaller values with increase in  WGW. Whereas variation of FSE for correlated two-photon (CTP) absorption is rather flat and remains high in a broader range of values of WGW than that for the STP absorption. Moreover the magnitude of FSE for CTP transition is higher than that for  the  STP transition in all the figures. 
 
(3) From these figures it is found that the fidelity of spatial Entanglement for both type of transitions (STP and CTP) depends on the photon frequency. The highest values of fidelity and the spread of the range of the higher values of fidelity increases with decrease of photon frequency of transition. For smaller values of photon frequency, molecular ion is excited nearer to the dissociation threshold. Hence the kinetic energy of dissociated atom and ion will be smaller than that for larger photon frequency, the recoil energy will also be smaller and the outgoing wavepacket spreads more slowly than that for higher kinetic energy of ion and atom \cite{Reid2014, Picon2011}. 
Hence it is easier to maintain coherence for a longer time which may preserve strong position correlations between ion and atom. However decoherence effect may affect the entanglement for smaller frequency dissociation and hence interplay of these effects will determine the optimal frequency to achieve maximum entanglement. 
  
  (4) It is also found that the difference in highest values of fidelity for STP and CTP dissociations decreases with decreases in photon frequency, but the highest value of fidelity for STP dissociation remains smaller than those for CTP absorption. Although the range of WGW for which FSE for STP dissociation remains high increases with decrease in photon frequency, it remains smaller than that  for CTP dissociation. This is because of the fact that when the molecular ion interacts with the nonlocal electromagnetic field, coherence between two photo-absorption processes is transmitted to the atom and ion through dissociation leading to stronger spatial entanglement than that for STP dissociation of molecular ion. 

\subsection{Dependence on time:}

To explore the dependence of fidelity of entanglement on the time at which fidelity was calculated, we have plotted FSE vs WGW for three values of time t=3000 a.u., 5000 a.u. and 6500 a.u. for each value of frequency in figures 5, 6 and 7 respectively. In each figure FSE for both type of dissociations CTP and STP has been shown for comparison.  
  
\begin{figure}
\centering
\includegraphics[width=0.65\textwidth]{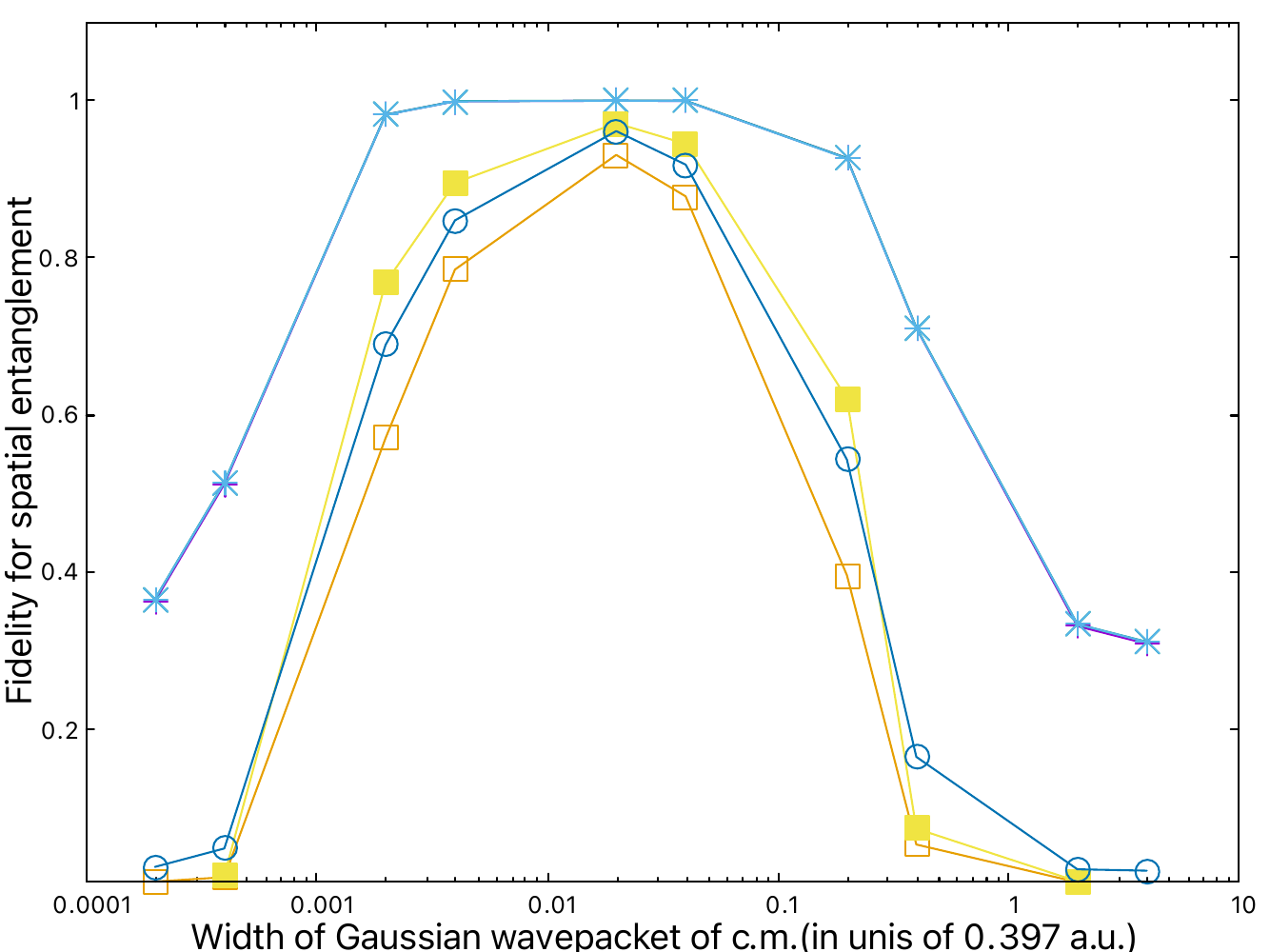}

 {Fig5: Fidelity for spatial entanglement between dissociated atom and ion from (i) sequential two photon dissociation and (ii) correlated two photon dissociation of ${H_2}^+$ ion as a function of initial width of the gaussian wavepacket of centre of mass at three values of time=3000 a.u., 5000 a.u. and 6500 a.u., for photon frequency = 0.368267 a.u. 
(i) For sequential two photon dissociation: 
Open square - time = 3000 a.u., 
Filled square - time = 5000 a.u., 
Open circle - time = 6500 a.u.
(ii) For correlated two photon dissociation:
Plus - time = 3000 a.u., 
Cross - time =5000 a.u.,
Star - time = 6500 a.u.}
\end{figure}

Comparing Figures 5,6 and 7, explicit dependence of FSE on choice of time can be obtained. It is found that in STP transition, for highest value of  photon frequency (0.36287 a.u.) (Fig. 5) the values of FSE and the range of WGW for which fidelity remains high change with increase in time. However for smaller values of photon frequencies (Figs. 6 and 7) the highest value of fidelity and the range of WGW for which fidelity remains high coincides for larger values of time (5000 a.u. and 6500 a.u.).  Which shows if the time at which the fidelity is calculated is large enough the fidelity of spatial entanglement saturates and hence the dependence on time becomes insignificant.

\begin{figure}
\centering
\includegraphics[width=.65\textwidth]{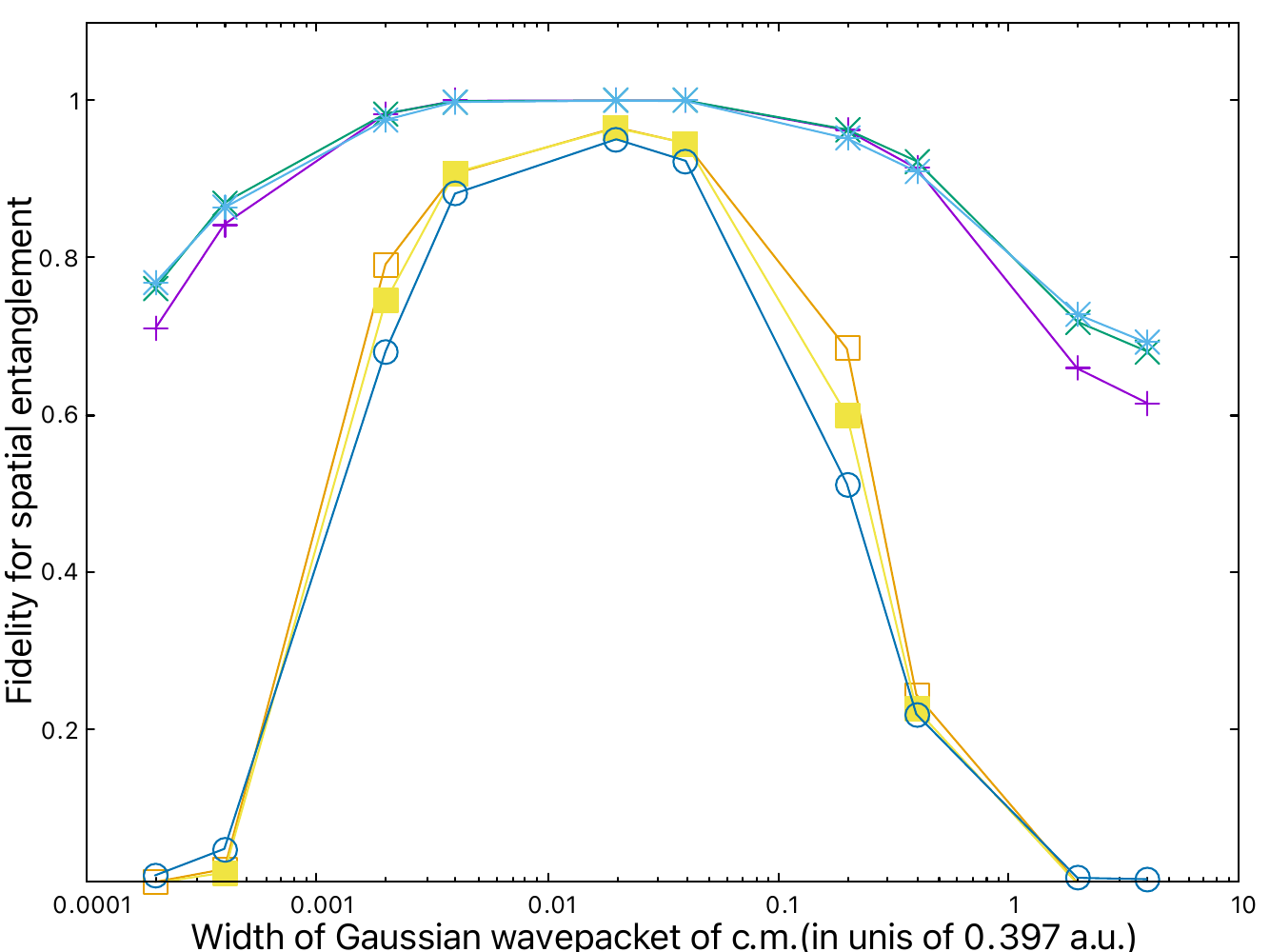}

 {Fig6: Fidelity for spatial entanglement between dissociated atom and ion from (i) sequential two photon dissociation and (ii) correlated two photon dissociation of ${H_2}^+$ ion as a function of initial width of the gaussian wavepacket of centre of mass at three values of time=3000 a.u., 5000 a.u. and 6500 a.u., for photon frequency=0.3059225 a.u.
Both for (i) sequential two photon dissociation and
(ii) correlated two photon dissociation: Symbols and the corresponding times  are the same 
as in fig5.}
\end{figure}

 The dependence of fidelity  for CTP on time is insignificant for all the photon frequencies chosen. This infers that when the correlation between ion and atom becomes strong the highest values of fidelity of spatial entanglement saturates even for smaller values of time and this is the case when molecular ion interacts with nonlocal mode of 
 electromagnetic field. This feature of saturation of fidelity with time at which fidelity is calculated infers  the robustness of the processes considered here.

\begin{figure}
\centering
\includegraphics[width=.65\textwidth]{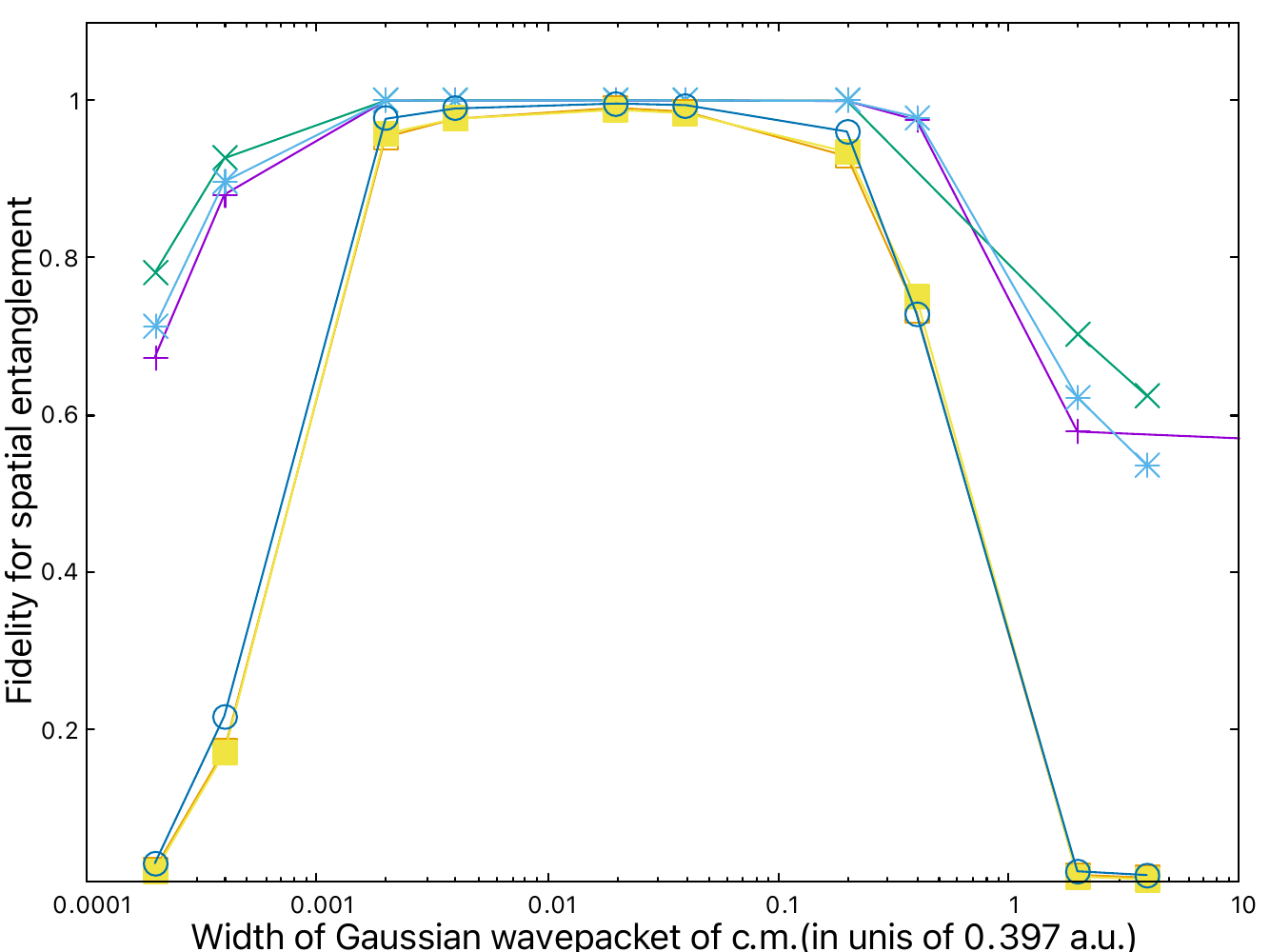}

{Fig7: Fidelity for spatial entanglement between dissociated atom and ion from (i) sequential two photon dissociation and (ii) correlated two photon dissociation of ${H_2}^+$ ion as a function of initial width of the gaussian wavepacket of centre of mass at three values of time=3000 a.u., 5000 a.u. and 6500 a.u., for photon frequency = 0.279422 a.u.
Both for (i) sequential two photon dissociation and 
(ii) correlated two photon dissociation: Symbols and the times  are the same as in fig5}
\end{figure}

   From the data for fidelity it is found that the highest value of fidelity obtained for CTP transition is 99.99 percent for the lowest frequency and at highest value of time. It remains the same at three consecutive values of WGW. Moreover  it remains high (within .02-.03 percent) in a broad range of WGW. For higher values of frequency and lower values of time fidelity remains high within .03-.05 percent of the highest value.  Whereas that for STP transition the highest value is 99.60 percent at a single value of WGW only for  lowest value of frequency and highest value of time. This is obvious from the figures presented here. Thus it is easier to achieve highest value of FSE and very high values of fidelity in a broad range of WGW, in CTP dissociation.
   
  It is to be mentioned here that for highest value of photon energy 0.368 a.u., the velocity of atom will be $1.124 \times {10^6} m/s$. Therefore atom will traverse a distance of $1\mu m$ in 0.889 picosecond. Present calculation is robust, since the values of fidelity both for CTP and STP dissociation saturate with time and that for CTP transition saturation occurs within a very short period of time 3000 a.u. Moreover, since for CTP dissociation, the highest value of fidelity is 99.99 percent and it remains high for a broad range of WGW,  this can facilitate long distance measurement of spatial entanglement of atom and ion, in this scheme.

\section{Conclusion}
In this work we calculated fidelity for spatial entanglement of dissociated H-atom and proton for sequential two photon absorption from ground $1s\sigma_g$ state to the dissociating $2s\sigma_g$ state via $2p\sigma_u$  state of ${H_2}^+$ ion  and for  simultaneous absorption of two photons, correlated in phase, leading to excitation directly from the ground $1s\sigma_g$ state to the dissociating $2s\sigma_g$ state of the molecular ion. Wavepacket formalism has been used to generate the wavepacket on final dissociating state and  the potential energy curves for this calculation was obtained from literature. The dipole transition moments required for sequential two-photon absorption were obtained from literature, whereas those for correlated two photon transition were derived analytically. 

Here we have extended our previous work \cite{TKRD1998} on interaction of atom with nonlocal mode of electromagnetic field to study the fidelity of spatial entanglement of dissociated atom and ion due to the interaction of molecular ion  with nonlocal mode of electromagnetic field. In this model, with increase in laser intensity photon flux in the interaction region will be high leading to simultaneous absorption of two phase correlated photons by the molecular ion within a very short time ( $\delta {t} << {1\over {\omega}}$, where $\omega$ is the photon frequency). Hence the correlation between two photo-absorption processes is transmitted to the dissociated atom and ion, increasing the correlation between them. This will lead to enhancement in the fidelity of spatial entanglement between dissociated ion and atom. By comparing these results with those from sequential two photon dissociation, it is shown that fidelity from CTP dissociation is always higher and remains high in a broader range of intial width of the gaussian wavepacket (WGW) of c.m. than those from STP dissociation. 

It is shown that the fidelity of spatial entanglement can be controlled by varying the frequency of photon, the lower the frequency, higher is the fidelity. Study on the dependence on time at which fidelity is calculated shows fidelity in both type of dissociation saturates with increase in time and for CTP absorption it saturates within a very short period of time. Hence this process of achieving high fidelity is robust and it is much more favorable for correlated two photon dissociation. 

A scheme for detection of spatial entanglement has been suggested which is similar to the integrated atom optics, where one of the arms will be replaced by a channel for detection of ion, i.e. it will contain ion waveguide, phase shifter, beam splitter and trapped ion detector.
 We have also shown that there
is a scope of long distance detection of spatial entanglement between dissociated atom and ion.

\section {Appendix}
For $H_{2}^+$ ion the co-ordinate frame is chosen such that atom 'H' is at the centre of the coordinate frame and the internuclear distance R is along the z-axis (see Fig1). The coordinate 'r' connecting 'H' and the electron is $\vec {r}=\vec {r_a}$. Hence $\vec {r_b}= \vec {r} - \vec{R}$ or $r_b= |\vec {r}-\vec {R}|$ and $|r_b|^2 = |r|^2+|R|^2- 2 |r||R| cos\theta$ where
$\theta$ is the angle between $\vec{r}$ and $\vec{R}$. Therefore electronic radial wave function for the ground ($1s\sigma_g$) and excited ($2s\sigma_g$) states are given as:

$$ R_i(r';R') = {1 \over {\sqrt{\pi a^3}}} {[\exp-{r'/a} + \exp{-{{\sqrt{|r'|^2+|R'|^2-2r'R'cos\theta}}\over {a}}]}} $$

and
$$R_f(r;R)= {1\over {4\sqrt{2\pi}}} [{{\exp-{r\over{2a}} (2-{r/a})}}] +  
[\exp{-{{\sqrt{|r|^2+|R|^2-2rRcos\theta}}\over {2a}}} $$
$$ \times{{{(2-{{\sqrt{|r|^2+|R|^2-2rRcos\theta}}\over {a}})}]}} $$

respectively.

Using these two wave functions the radial integral $\textbf{R}$ (13) was evaluated analytically. Hence $T_{if} (R,R')$ was calculated considering $l_f=l_i=0$ and $m_f=m_i=0$. Using analytically calculated electronic dipole transition moment, the nuclear wavepacket on the final dissociating state has been calculated and using this wavepacket fidelity is calculated as described in section 2-'Theory'.

\textbf{Acknowledgement}:

I Thank B. Deb for his interest in this work and for useful discussions during the work.

\end{document}